\documentstyle[epsf,aps]{revtex}
\begin{document}
\bibliographystyle{osa}
\title{In--out intermittency in PDE and ODE models}
\author{Eurico Covas${}^{1}$\thanks{Web: http://www.eurico.web.com},
Reza Tavakol${}^{1}$\thanks{Email: reza@maths.qmw.ac.uk},
Peter Ashwin${}^{2}$\thanks{Email: P.Ashwin@ex.ac.uk},
Andrew Tworkowski${}^{3}$\thanks{Email: ast@maths.qmw.ac.uk} \&
John M.\ Brooke${}^{4}$\thanks{Email: J.M.Brooke@mcc.ac.uk}}
\address{
1. Astronomy Unit, Mathematical Sciences, Queen Mary, University of London,
Mile End Road, London, United Kingdom\\
2. School of Mathematical Sciences, Laver Building, University of Exeter,
Exeter EX4 4QE, United Kingdom\\
3. Mathematics Research Centre, Mathematical Sciences, Queen Mary, University of London,
Mile End Road, London, United Kingdom\\
4.CSAR HPC Support Group Manchester, Computing Centre, Oxford Road,
Manchester, United Kingdom}
\date{\today}
\maketitle
\begin{abstract}
We find concrete evidence for a recently discovered
form of intermittency, referred to as in--out intermittency,
in both PDE and ODE models of mean field dynamos.
This type of intermittency (introduced in \cite{ashwin99})
occurs in systems with invariant submanifolds and, as 
opposed to on--off intermittency which can also occur in skew product systems,
it requires an absence of skew product structure. By this we mean that
the dynamics on the attractor intermittent to the invariant manifold 
cannot be expressed simply as the dynamics on the invariant subspace
forcing the transverse dynamics; the transverse dynamics will alter that
tangential to the invariant subspace when one is far enough away from the
invariant manifold.

Since general systems with invariant submanifolds are not likely to
have skew product structure, this type of behaviour may be of physical
relevance in a variety of dynamical settings.

The models employed here to demonstrate in--out intermittency are
axisymmetric mean--field dynamo models
which are often used to study the observed
large scale magnetic variability in the Sun and solar-type stars.
The occurrence of this type of intermittency in such
models may be of interest in understanding some
aspects of such variabilities.
\end{abstract}
\vspace{0.5cm}
{\bf 
Dynamical systems that possess symmetries (and hence invariant
submanifolds embedded in their state spaces) are of interest
in a variety of settings. In many simplified models
such dynamical systems have skew product structure. For an ODE model,
if $(x,y)$ parameterizes a phase space with an invariant manifold $y=0$,
we say the system has skew product structure if $\dot{x}=f(x)$ 
and $\dot{y}=g(x,y)$, namely if the dynamics of $x$ is independent of $y$. 
A great deal of effort has gone into the study of such skew product systems 
with invariant manifolds, and these have thrown up a number of 
new and interesting phenomena.

In general, however, one would expect dynamical systems not to 
have skew product structure unless extra structure is present (for example if
the transverse dynamics is always forced by the tangential dynamics).  
In the absence of such extra structure it is therefore
interesting to see what new types of dynamics can appear in systems 
with invariant submanifolds.
One such novel type of dynamical behaviour, in--out intermittency,
is discussed and analysed in detail in \cite{ashwin99} using a simple 
two-dimensional mapping.  An important feature of this type of 
intermittency is that, as opposed to on--off intermittency, it {\em requires}
the absence of a skew product structure.

In this paper we find concrete evidence for the occurrence of 
in--out intermittency in both PDE and ODE models both
in terms of phase--space and also statistically. The models considered are
examples of axisymmetric mean--field dynamo models
which are often used in order to study the observed
large scale magnetic variability in the Sun and solar-type stars.
In addition to providing examples of in--out intermittency in
PDE models, the occurrence of this type of intermittency in such
models may be of interest in understanding some
aspects of solar and stellar variabilities.
}
\section{Introduction}
Many systems of physical interest possess symmetries
which in turn induce invariant submanifolds in their state spaces.
A great deal of effort has gone into the study of the dynamics
and intermittent behaviour of such systems
near their invariant submanifolds
(see e.g.\ \cite{plattetal1993}).
A class of dynamical systems with invariant submanifolds
have been shown to be capable of producing a number of novel
modes of behaviour, including on--off intermittency,
which occurs as the result of an
instability of an attractor in an invariant submanifold 
\cite{plattetal1993}. It manifests itself as
an attractor whose trajectories get arbitrarily close to an 
attractor for the system in the invariant submanifold while 
intermittently making large deviations away. It can be 
modelled by a biased random walk of the
logarithmic distance from the invariant submanifold \cite{plattetal1993}.

Since the linearised behaviour near an invariant submanifold has 
a natural skew product structure (i.e.\ the linearised dynamics 
transverse to the invariant submanifold is forced by the dynamics 
within the submanifold) many such studies have tended to 
concentrate on systems that are of {\em skew product} type for simplicity,
although it should be stated that on-off intermittency can be found in
systems that do not have skew product structure.

Moreover, bifurcation problems in such settings have tended to 
concentrate on {\em normal} parameters, i.e.\ parameters that 
vary the global dynamics without changing the dynamics within an 
invariant submanifold. In general, dynamical systems are not skew products 
over the dynamics within any invariant subspace, and moreover they do {\em not}
possess normal parameters\cite{invsubspace}. 

The authors \cite{covas97d,ashwin99} have recently shown that dropping these
assumptions can lead to the presence of a number of novel
types of dynamical behaviour, including a new type of
intermittency, referred to as {\em in--out intermittency}.
The presence of this type of intermittency has also been
found in different distinct nonlinear dynamical systems 
\cite{ashwin99,zhangetal2000}. Furthermore, there have been 
interesting developments concerning the study of other
phenomena -- e.g.\ {\em riddling} -- in these more general settings
\cite{laietal1999}.

To characterise in--out intermittency, it is best to contrast it with on--off
intermittency, as they both can occur in systems with invariant submanifolds.
To begin with, it is useful to bear in mind that
even though on--off intermittency can occur
in non--skew product settings, all its necessary ingredients
can be satisfied in skew product settings.
In--out intermittency, on the other hand, requires
the {\em absence} of skew product structure for its existence.

Briefly, we say that an attractor $A$ exhibits {\em in--out intermittency}
to the invariant submanifold $M_I$, if the following are true \cite{ashwin99}:

\begin{enumerate}
\item
The intersection $A_0=A\cap M_I$ is not necessarily a
minimal attractor, i.e.\ there can be proper subsets of $A_0$ that are
attractors (for on--off intermittency $A_0$ is assumed to be minimal). This
means that there can be different invariant sets in $A_0$ associated with
attraction and repulsion transverse to $A_0$, hence the name in--out.
These growing and decaying phases come about through different mechanisms
within $M_I$. If the system has a skew--product structure, in--out 
intermittency reduces to on--off intermittency \cite{ashwin99}.
Fig.\ \ref{in-out-trajectory} shows a schematic representation of a typical
trajectory for an in--out process near $M_I$.

\begin{figure}
\centerline{\def\epsfsize#1#2{0.54#1}\epsffile{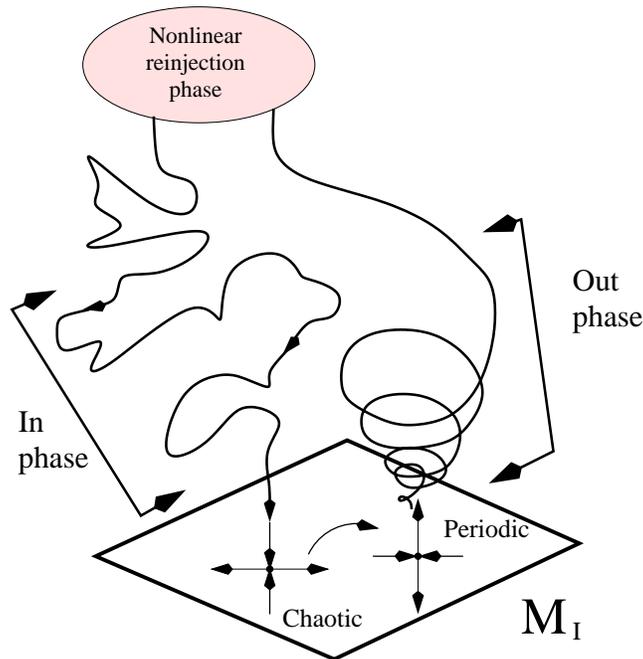}}
\caption{\label{in-out-trajectory} Typical trajectory of an in--out
intermittent solution close to the invariant submanifold $M_I$, with the
two components, the ``in'' phase and the ``out'' phase. In the invariant
submanifold $M_I$ we may have
two or more invariant sets, one of which is
transversely stable and chaotic but non--attracting in $M_I$ and another
which is transversely unstable and is a periodic attractor in $M_I$.
The injection mechanism, the ``in'' phase, is quite irregular and
can be modelled by a random walk towards $M_I$, while the expelling
mechanism, the ``out'' phase, can be modelled by a growing exponential spiral
away from $M_I$. Note that the invariant sets in $M_I$ are represented as
points only for clarity.
}
\end{figure}

\item
The minimal attractors in the invariant submanifold are not necessarily
chaotic (as for on-off intermittency); they are very
frequently periodic or equilibria. Furthermore, the trajectory remains
close to one of these attractors during the moving away or ``out''
phases, with the important consequence that during these ``out'' phases
the trajectory can shadow a periodic orbit, for example, while drifting
away from $M_I$ at an exponential rate \cite{ashwin99} (see also
\cite{brooke97}).

\item
The asymptotic scaling of the probability distribution of the duration of
laminar phases in the in--out case can have two contributions:
\begin{equation}
P_{n}\sim \overbrace{\underbrace{\alpha n^{-3/2}e^{(-\beta n)}}_{\rm
on-off} + \gamma e^{(-\delta n)}}^{\rm in-out}= I_1+I_2,\label{scaling}
\end{equation}
where $\alpha>0$, $\delta>\beta>0$ and $\gamma>0$ are positive real
constants depending on the bias of the random walk modelling the ``in''
phase and the probability of leaking into the deterministic ``out''
phase (see \cite{ashwin99} for details). The term $I_1$ corresponds to
biased on--off intermittency, while the extra term $I_2$ can cause
an identifiable shoulder to develop at large laminar sizes $n$ which can
help to statistically distinguish in--out from on--off intermittency.

\end{enumerate}
The authors in \cite{ashwin99} were motivated by a numerical exploration
of a two dimensional map and explored the statistics by means of a Markov
chain model. Our aims in this paper are twofold. Firstly, we demonstrate
the occurrence of in--out intermittency in dynamical systems generated by
ordinary differential equations (ODE) as
well as by partial differential equations (PDE).  The latter are especially of
interest, since they are in principle infinite dimensional and also because
few examples of intermittent behaviour and their scalings have been shown 
concretely to occur in
such models (see e.g.\ \cite{scalings}).  Secondly, by choosing as our models
the mean--field dynamo models \cite{mf-models}, 
the occurrence of this type of intermittency could be of interest
in understanding certain features of solar and stellar variability, 
and in particular we expect that due to its generic features, it
may well appear in more detailed and accurate models of solar and stellar
variability.
\section{In--out intermittency in Mean--Field dynamo models}
Mean-field dynamo models have been employed extensively in order to
study various aspects of the dynamics of solar, stellar and galactic
dynamos (e.g.\ \cite{brandenburg89,applications}).  Their rather
idealised nature has been criticized by a number of authors (see e.g.\
\cite{critique}).  However, such models are thought to capture some of the
essential physics of the turbulent processes and reproduce many
important dynamical and statistical features of the full three
dimensional magneto-hydrodynamical models (see e.g.\ \cite{support} and
also \cite{theoretical_good}).

The standard mean--field dynamo equation is given by
\begin{equation} \label{dynamo}
\frac{\partial {\bf B}}{\partial t}=
\nabla \times \left( {\bf u} \times {\bf B} + \alpha {\bf B} - \eta_t
\nabla \times {\bf B} \right),
\end{equation}
where ${\bf B}$ and ${\bf u}$ are the mean magnetic field and mean
velocity respectively and the turbulent magnetic diffusivity $\eta_t$
and the coefficient $\alpha$ arise from the correlation of small scale
turbulent velocities and magnetic fields \cite{mf-models}.

In axisymmetric
geometry, equation (\ref{dynamo}) is solved by splitting the magnetic field
into poloidal and toroidal components, ${\bf B} = {\bf B_{p}} + {\bf
B_{\phi}}$, and expressing these components in terms of scalar field
functions
$$
{\bf B_{p}} = {\bf \nabla}\times A(r,\theta,t){\bf \hat{\phi}},~~
{\bf B}_{\phi} = B(r,\theta,t){\bf \hat{\phi}},
$$
in spherical polar coordinates $(r, \theta, \phi)$.
Equation (\ref{dynamo}) can then be expressed in terms of equations for
the scalars $A$ and $B$,
\begin{eqnarray}
\label{alpha2omega}
\frac{\partial A}{\partial t} &=& \alpha B + \eta_t \left(
\nabla^2
-\frac{1}{r^2\sin^2\theta}\right) A,\\
\nonumber
\frac{\partial B}{\partial t} &= & r  \sin \theta ({\bf \nabla}\times A {\bf \hat{\phi}})
  \cdot{\bf \nabla}\omega\\\nonumber & &-\frac{1}{r\sin\theta}{\bf \nabla}
\alpha\cdot{\bf \nabla}(A r \sin\theta)\\\nonumber & & - \alpha \left( \nabla^2 -
\frac{1}{r^2\sin^2\theta}\right) A
\\\nonumber &&
+ \eta_t \left( \nabla^2 - \frac{1}{r^2\sin^2\theta}\right) B,
\nonumber
\end{eqnarray}
where ${\bf \nabla}\cdot A =
0$ and we consider a purely rotational velocity ${\bf u}= \omega_0 r^2 \sin\theta\hat{\phi}$.
Nondimensionalisation of these equations in terms of a length $R$ and a time
$R^2/\eta_t$ produces the convective and
rotational magnetic Reynolds numbers $C_\alpha=\alpha_0 R/\eta_t$ and
$C_\omega=\omega_0 R^2/\eta_t$, where $\alpha_0$ and $\omega_0$ are
typical values of $\alpha$ and $| \omega|$.

Solutions to these equations are often considered in
the $\alpha \omega$ limit where the terms in $\alpha$
can be ignored in the equation for $B$, giving a single dynamo parameter
$D = C_\alpha C_\omega$ on rescaling. This reflects the fact that, in stellar convective
zones, rotational shear produces toroidal flux much more effectively than the processes
represented by the $\alpha$ terms, whilst in the full equations (the so called $\alpha^2 \omega$ limit)
we retain both $C_\alpha$ and $C_\omega$ as two control parameters.

The equation (\ref{dynamo}) gives a kinematic dynamo, since the velocity
field ${\bf u}$ is prescribed. As this equation stands there is no mechanism
to limit the growth of the magnetic field a nonlinear saturation
mechanism is usually supplied by making $\alpha$ depend on ${\bf B}$. This can
be done by supplying a closed functional form representing a fixed
approximation of the nonlinear effect (c.f.\
\cite{kitchatinov,tworkowskietal1998a}), or more dynamically, by
supplying an auxiliary equation for $\alpha$ (c.f.\ \cite{covasetal1998a}
and references therein).

\begin{figure}
\centerline{\def\epsfsize#1#2{0.47#1}\epsffile{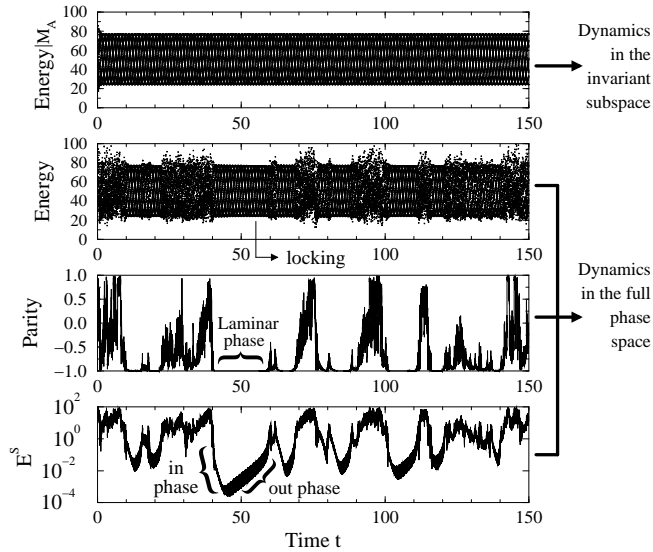}}
\caption{ \label{inoutpde1}
In--out intermittency in the axisymmetric PDE mean--field dynamo model
(\ref{alpha2omega}).
The parameters used were $r_0=0.4$, $C_{\alpha}=1.942$, $C_{\omega}=-10^{5}$,
$f=0.0$, with the usual algebraic form of
$\alpha=\alpha_0/(1+{\bf B}^2)$ (see {\protect \cite{tworkowskietal1998a}} for details of
the parameters). To visually enhance the periodic locking we have time sampled the
series in the two upper panels.
}
\end{figure}

\begin{figure}
\centerline{\def\epsfsize#1#2{0.47#1}\epsffile{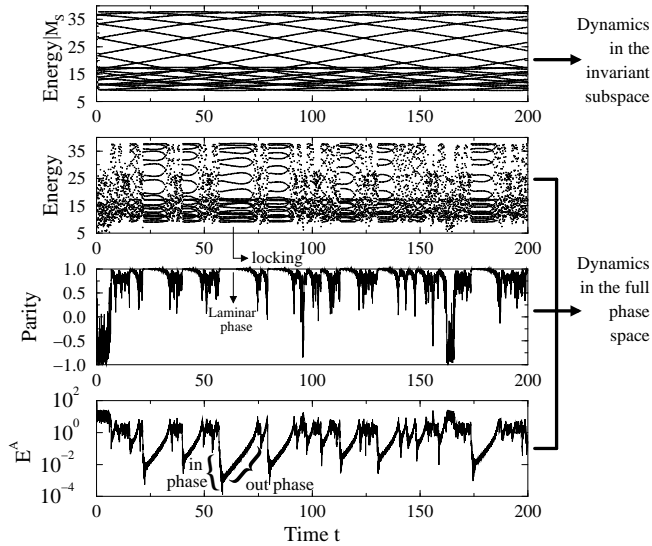}}
\caption{ \label{inoutpde2}
In--out intermittency in the axisymmetric PDE mean--field dynamo model (\ref{alpha2omega}).
The parameters used were $r_0=0.4$, $C_{\alpha}=1.5$, $C_{\omega}=-10^{5}$,
$f=0.7$, together with an algebraic form of $\alpha$ due to Kitchatinov
{\protect \cite{kitchatinov}}. The two upper panels are shown as in Fig.\
\ref{inoutpde1}.
}
\end{figure}

In the following we consider two cases arising from two separate studies
\cite{covas97d,tworkowskietal1998a}: the above PDE model in the
$\alpha^2 \omega$ limit with two different algebraic forms for
$\alpha(\bf{B})$ (c.f.\ \cite{kitchatinov,tworkowskietal1998a} and
Figs.\ \ref{inoutpde1} and \ref{inoutpde2} captions) as well as a
finite order truncation of it in the $\alpha \omega$ limit but with a
time dependent form of the $\alpha$ effect in one spatial dimension
(this can be obtained by averaging (\ref{dynamo}) over $r$) and using a
spectral expansion \cite{covas97d}. This ODE model possesses a second (alongside $D$)
control parameter, the magnetic Prandtl number $\nu = \nu_{t} /
\eta_{t}$, where $\nu_{t}$ is the turbulent kinematic viscosity, which
arises from the time dependent equation for $\alpha$. This model
is given by

\begin{eqnarray}
\frac{dA_i}{dt}&=&-i^2 A_i+\frac{D}{2}(B_{i-1}+B_{i+1})\label{truncated}
\\&&+
\sum_{j=1}^{N}\sum_{k=1}^{N} F_{ijk} B_j C_k,\nonumber\\
\frac{dB_i}{dt}&=&-i^2 B_i+\sum_{j=1}^{N} G_{ij}A_j, \nonumber\\
\frac{dC_i}{dt}&=&-\nu i^2 C_i
-\sum_{j=1}^{N}\sum_{k=1}^{N} H_{ijk} A_j B_k,\nonumber
\end{eqnarray}
where $A_i$, $B_i$ and $C_i$ are spatially independent coefficients of the spectral
expansions of the scalar fields $A$, $B$ and $\alpha$ respectively, $F$,
$H$ and $G$ are coefficients expressible in terms of $i,j$ and $k$, $N$ is
the truncation order and $D$ and $\nu$ are the parameters
defined above.
The detailed derivation of these equations together with a phenomenological
study of their dynamics
is given in \cite{covas97d}.

We note that the main ingredients necessary for the occurrence of in--out
intermittency are present in both these models. Both are axisymmetric and
possess invariant submanifolds. More precisely,
the truncated model (\ref{truncated}) with $N$=4
is a 12--dimensional system of ODEs with two 6--dimensional symmetric and
antisymmetric invariant submanifolds given by
\begin{equation}\label{ms}
M_S=\{0,B_1,0,A_2,0,C_2,0,B_3,0,A_4,0,C_4\},
\end{equation}
\begin{equation}\label{ma}
M_A=\{A_1,0,0,0,B_2,C_2,A_3,0,0,0,B_4,C_4\},
\end{equation}
respectively. Similarly the
PDE model (\ref{alpha2omega}) possesses two invariant submanifolds,
the antisymmetric and symmetric invariant
submanifolds which are given by
\begin{equation}
\label{submanifolds}
{\rm M}_A\ :\ A(\theta)=A(-\theta),  \quad  B(\theta)=-B(-\theta),
\end{equation}
\begin{equation}
\label{submanifolds2}
{\rm M}_S\ :\ A(\theta)=A(-\theta),  \quad  B(\theta)= B(-\theta),
\end{equation}
respectively, where $\theta$ is
the latitude.

If one separates the poloidal and toroidal scalar field
components into symmetric and antisymmetric parts then the dynamic
evolution for the symmetric (antisymmetric) components has contributions
from antisymmetric (symmetric) counterparts. This means that
these equations are of non--skew product type. 
For the ODE system (\ref{truncated})
this can be readily seen by noting that the evolution equation for each
component in $M_S$ ($M_A$) contains components from $M_A$ ($M_S$).  For
the PDE models, we first note that ${\bf u}$ in equation (\ref{dynamo}) is
not dynamical: it is prescribed and therefore can be viewed as a part of
the initial conditions. The non-skew product nature of the PDE models
follows in a similar way to the ODE models, bearing in mind the form of
the equation (\ref{dynamo}) and those of the invariant submanifolds
(\ref{submanifolds}) and (\ref{submanifolds2}).

Finally the control parameters $D$ and $\nu$ appearing in the ODE model
(\ref{truncated}) are generically non--normal as they enter the equations
for $A_i$ and $C_i$ for all $i$.  Similarly this is also true for the control
parameters $C_\alpha$ and $C_\omega$ in the case of
the nondimensionalised version of the PDE equations (\ref{alpha2omega}).

In this way, both models possess all the necessary ingredients for the
occurrence of in--out intermittency.

\begin{figure}
\centerline{\def\epsfsize#1#2{0.47#1}\epsffile{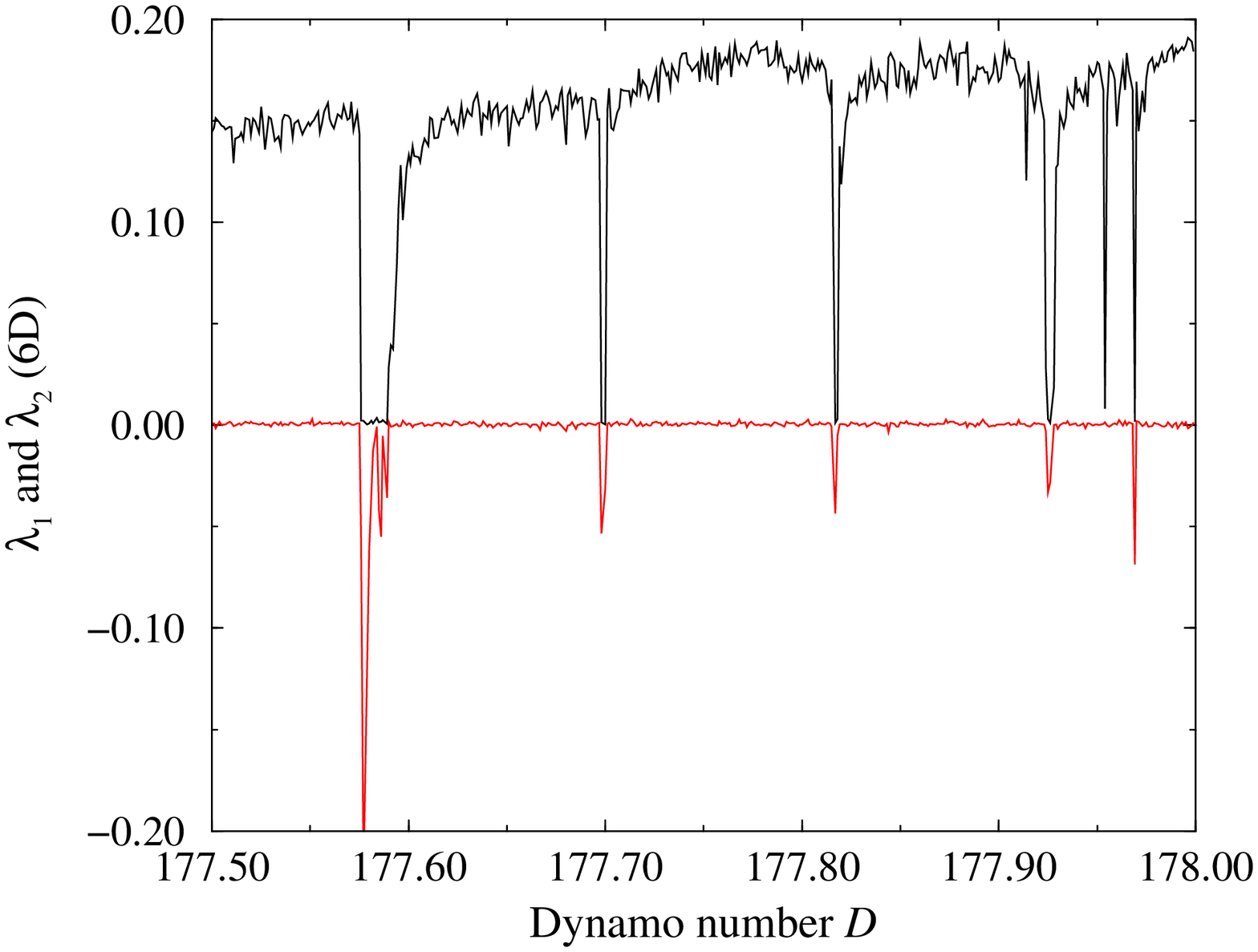}}
\caption{\label{lyapunov}
First and second Lyapunov exponents for the attractor of the
ODE model (\ref{truncated}) restricted to $M_A$ with $N$=4 and
$\nu=0.47$, clearly showing the presence of {\em windows of periodicity}.
Note that there is always one null Lyapunov exponent due to
the system of equations (\ref{truncated}) being autonomous.
}
\end{figure}

Given that the ODE models are more transparent, we first  demonstrate the
presence of in--out intermittency in the truncated system
(\ref{truncated}) with $N$=4. For this model, in--out intermittency occurs
for parameter values for which the system of equations (\ref{truncated})
restricted to $M_A$ is within a {\em window of periodicity} (c.f.\
\cite{barretoetal1997}).
Fig.\
\ref{lyapunov} depicts the presence of such windows for the
ODE model (\ref{truncated}).
The presence of such windows is supported by a conjecture of
Barreto {\em et al.} \cite{barretoetal1997}, according to which
for chaotic systems with $k$ positive Lyapunov exponents and $m$
control parameters, with $m\ge k$, there is a dense
set of nearby parameter values at which the attractors are
periodic.
This implies that for our system (\ref{truncated}),
for each parameter value at which there is
a chaotic attractor in $M_A$ there are parameter windows
arbitrarily close for which the attractor is periodic.

\begin{figure}
\centerline{\def\epsfsize#1#2{0.47#1}\epsffile{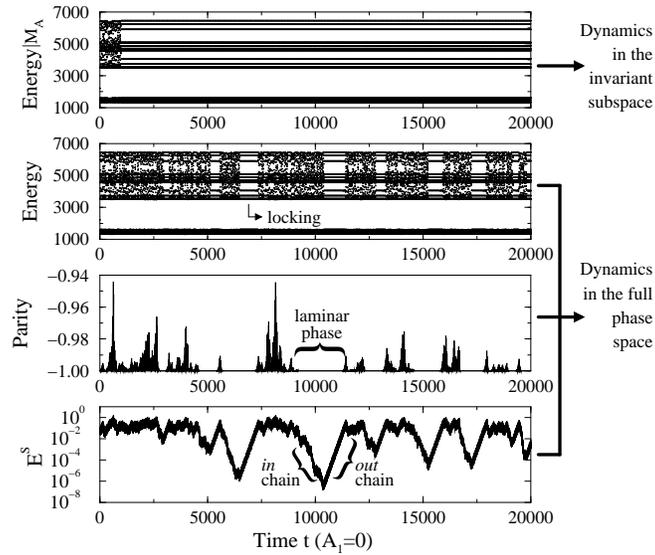}}
\caption{\label{inoutode}
In--out intermittency in the ODE model (\ref{truncated}) with $N$=4 and
parameter values $D=177.700196$ and $\nu=0.47$. The energy and parity are given by
$E=E^A+E^S$ and $P=(E^S-E^A)/E$ respectively, where $E^A$ and $E^S$ are the
antisymmetric and symmetric parts of the magnetic field energy with respect to
the equator (``antisymmetric'' ($P=-1$) and ``symmetric'' ($P=+1$)).
The top panel shows the evolution of an initial condition in $M_A$ and the other
panels a nearby initial condition not in $M_A$. In these panels, we have taken
a Poincar\'e section at $A_1$=0 for clarity and
comparison.}
\end{figure}

Fig.\ \ref{inoutode} shows an example of
in--out intermittency in this system at parameter values $D=177.700196$ and
$\nu=0.47$.
We note that even though the interval reported here
over which in--out occurs is small, nevertheless there
are likely to be other intervals (according to the conjecture
of Barreto {\em et al.}, possibly an infinite number of them) over which this happens.

The top panel shows the periodic orbit in the antisymmetric invariant
submanifold, $M_A$, which the projection of the trajectory of the full system
shadows clearly (second panel). This shadowing or intermittent periodic
locking of the tangential variables occurs within the laminar phases (third
panel) where there is a simultaneous exponential growth (hence the name
``out'' phase) of the amplitudes of the transverse variables through several
orders of magnitude (bottom panel). This last panel also shows the ``in''
phases, which can be modelled as a biased random walk taking the trajectory
into the invariant submanifold.

\begin{figure}
\centerline{\def\epsfsize#1#2{0.44#1}\epsffile{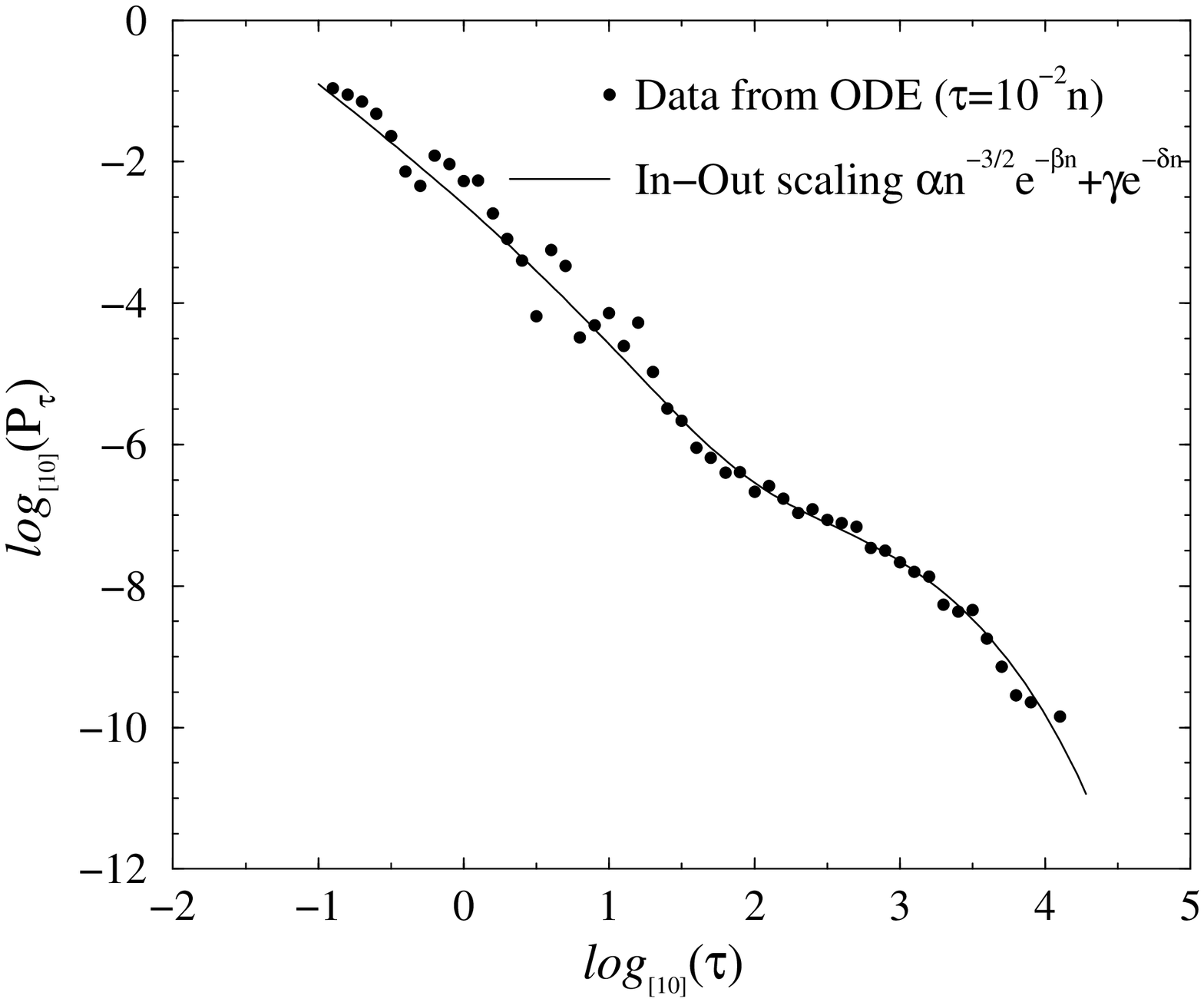}}
\caption{\label{scalingode}
Scaling of the probability distribution of the duration of laminar phases for
the twelve--dimensional ODE truncated model (\ref{truncated}) for the case considered in
Fig.\ \ref{inoutode}, using a time step $\tau=10^{-2}n$. The shoulder at large laminar
phases (which identifies the influence of $I_2$ and is a characteristic of
in--out intermittency) is easily discerned.}
\end{figure}

To substantiate this further, we
also calculated the scaling of the probability distribution of the
duration of laminar
phases and this is shown in Fig.\ \ref{scalingode}. This is compatible with
the predicted scaling (\ref{scaling}), possessing
both a $n^{-3/2}$ section, at small laminar phase sizes,
as well as a noticeable shoulder at higher laminar phase sizes,
the latter being a distinctive signature of the in-out
intermittency.


These signatures, namely the periodicity of the attractor of the system
restricted to the invariant submanifold, the periodic locking and the
exponential growth of the ``out'' phases together with the compatibility with the scaling
(\ref{scaling}) clearly show the occurrence of in--out intermittency in the
truncated ODE dynamo systems.

\begin{figure}
\centerline{\def\epsfsize#1#2{0.44#1}\epsffile{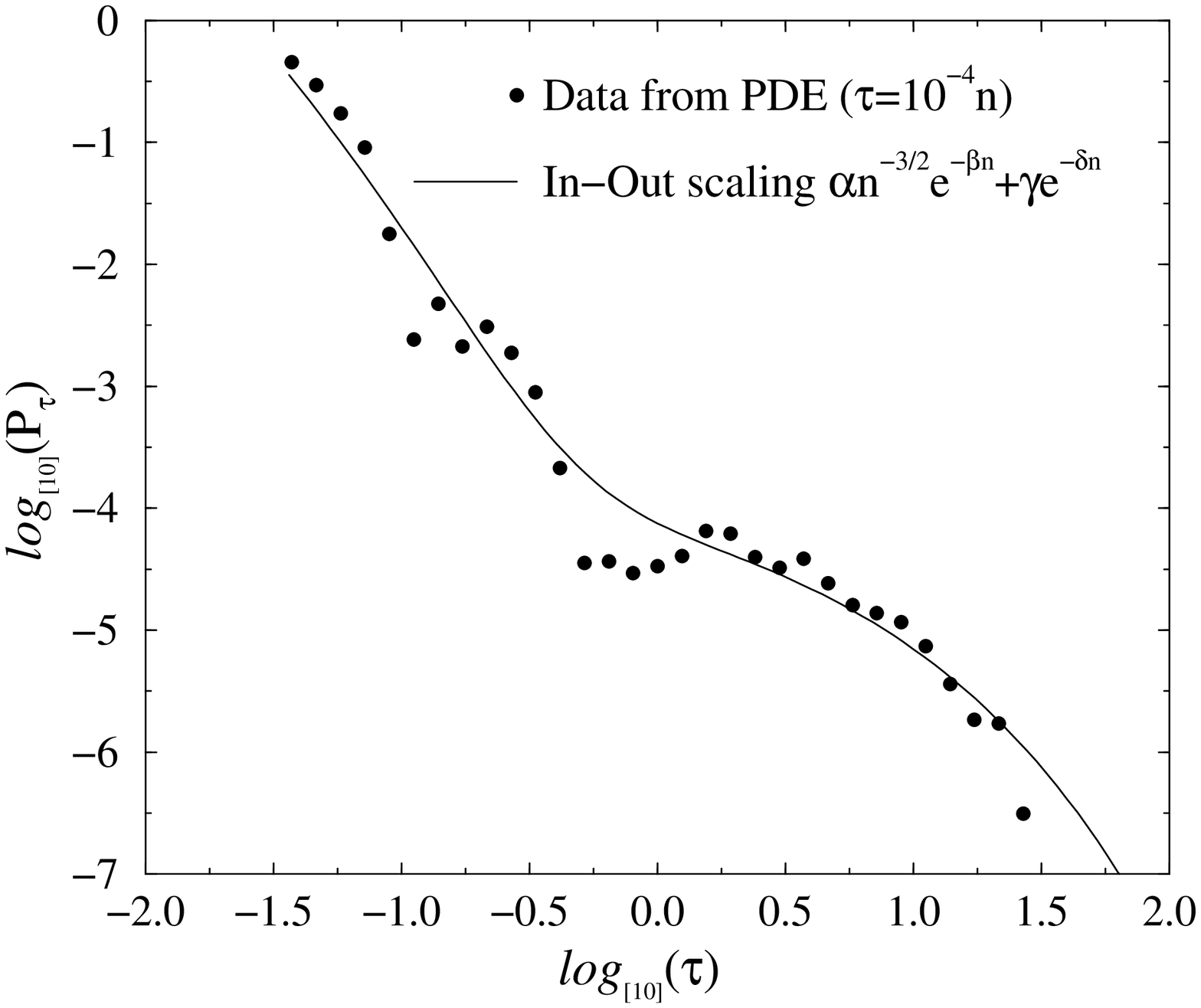}}
\caption{\label{scalingpde}
Scaling of the probability distribution of the duration of laminar phases for
the axisymmetric PDE mean--field dynamo model
(\ref{alpha2omega}) for the case considered in
Fig.\ \ref{inoutpde2}, using a time step $\tau=10^{-4}n$. The shoulder at large laminar
phases (which identifies the influence of $I_2$ and is characteristic of
in--out intermittency) is easily discerned.}
\end{figure}

To demonstrate the occurrence of in--out intermittency in the PDE case (which
as shown above also possesses all of the required ingredients), we integrated
equation (\ref{alpha2omega}), in parameter regions suggested by \cite{tworkowskietal1998a}, using
the code described in \cite{brandenburg89} and implemented by
\cite{tavakol95}.
Figs.\ \ref{inoutpde1} and \ref{inoutpde2} give examples of in--out
intermittency in these PDE models\cite{pdemodels}. As can be
seen, this behaviour can occur with the invariant submanifold being either
antisymmetric (Fig.\ \ref{inoutpde1}) or symmetric\cite{symmetry} (Fig.\
\ref{inoutpde2}). Again, in addition to the presence of periodic behaviour in
the system restricted to the invariant submanifold (top panels), these figures
clearly show the presence of locking during the ``out'' phases (second panels)
with an exponential growth of the energy of the transverse modes through
several orders of magnitude (bottom panels). This behaviour mirrors very
closely the truncated ODE model shown in Fig.\ \ref{inoutode} as well as that
expected to occur from the theory \cite{ashwin99}. To substantiate this
further, we again looked at the compatibility
of the scaling for the distribution of the laminar
phases with the theoretical scaling given by (\ref{scaling}).
Despite the greatly enhanced numerical cost of integrating the PDE
equations long enough to obtain convergence to the scaling law,
we have been able to establish agreement in this case
as can be seen in Fig.\ \ref{scalingpde}. Together, these signatures
clearly demonstrate the occurrence of in--out intermittency in these
 PDE dynamo models.

\section{Discussion}
By establishing the main ingredients necessary for the
occurrence of in--out intermittency as well as checking
the predicted corresponding phase space signatures and predicted scalings,
we have concretely demonstrated the occurrence of this
type of intermittency in
both ODE and PDE models. This type of intermittency requires
for its existence the non--skew product feature, the generality
of which makes the occurrence of this
type of intermittency of potential interest.

The models chosen here are mean--field dynamo models, which despite
their approximate nature are thought to capture many features
of magnetic activity in solar--type stars.  An important observed
feature of variabilities in solar-type stars is the presence of
dynamical behaviour with different statistics over different time
intervals due to the occurrence of the so called grand minima during
which the amplitude of the magnetic activity is greatly diminished.  A
number of scenarios have been suggested in order to explain these
phenomena (see e.g.\ \cite{explain}).  Within the
deterministic framework, intermittency \cite{proposal} (and multiple
intermittency \cite{multiple}) has been put forward as a
possible mechanism.  A number of studies have found
intermittent types of behaviour in such models (e.g.\
\cite{intermittent-type} and references therein).
The concrete demonstration of in--out as well as
other forms of intermittency are of potential 
importance in this regard as they demonstrate
the possible types of 
dynamical variability that can occur in such settings.


We thank Axel Brandenburg and David Moss for helpful conversations. EC
is supported by a PPARC postdoctoral fellowship, PA was partially
supported by EPSRC grant GR/K77365 and RT
benefited from PPARC UK Grant No.\ L39094.



\begin{thebibliography}{999}
\bibitem{symmetry}
We note that the dynamics typically select only one of the
invariant submanifolds (symmetric and antisymmetric) as containing
stable attractors. Each submanifold can have special significance in
accounting for observations of a particular star, like the Sun.
\end{thebibliography}
\end{document}